# The Lyman-α Sky Background as Observed by New Horizons

**G. Randall Gladstone,**[1,2] **Wayne R. Pryor,**[3] **S. Alan Stern,**[4] **Kimberly Ennico,**[5] **Catherine B. Olkin,**[4] **John R. Spencer,**[4] **Harold A. Weaver,**[6] **Leslie A. Young,**[4] **Fran Bagenal,**[7] **Andrew F. Cheng,**[6] **Nathaniel J. Cunningham,**[8] **Heather A Elliott,**[1,2] **Thomas K. Greathouse,**[1] **David P. Hinson,**[9] **Joshua A. Kammer,**[1] **Ivan R. Linscott,**[9] **Joel Wm. Parker,**[4] **Kurt D. Retherford,**[1,2] **Andrew J. Steffl,**[4] **Darrell F. Strobel,**[10] **Michael E. Summers,**[11] **Henry Throop,**[12] **Maarten H. Versteeg,**[1] **Michael W. Davis,**[1] **and the New Horizons Science Team**

[1]Southwest Research Institute, San Antonio, TX 78238.

[2]University of Texas at San Antonio, San Antonio, TX 78249.

[3]Central Arizona College, Coolidge, AZ 85122.

[4]Southwest Research Institute, Boulder, CO 80302.

[5]NASA Ames Research Center, Space Science Division, Moffett Field, CA 94035.

[6]The Johns Hopkins University Applied Physics Laboratory, Laurel, MD 20723.

[7]University of Colorado, Boulder, CO 80303.

[8]Nebraska Wesleyan University, Lincoln, NE 68504.

[9]Stanford University, Stanford, CA 94305.

[10]The Johns Hopkins University, Baltimore, MD 21218.

[11]George Mason University, Fairfax, VA 22030.

Corresponding author: first and last name (rgladstone@swri.edu)

**Key Points:**

- New Horizons Alice observations of interplanetary medium Lyman alpha are presented
- The observed brightness falloff with distance from the Sun matches well with Voyager results, and indicates a substantial external background contribution

**Abstract**

Recent observations of interplanetary medium (IPM) atomic hydrogen Lyman-α (Lyα) emission in the outer solar system, made with the Alice ultraviolet spectrograph on New Horizons (NH), are presented. The observations include regularly spaced great-circle scans of the sky and pointed observations near the downstream and upstream flow directions of interstellar H atoms. The NH Alice data agree very well with the much earlier Voyager UVS results, after these are reduced by a factor of 2.4 in brightness, in accordance with recent re-analyses. In particular, the falloff of IPM Lyα brightness in the upstream-looking direction as a function of spacecraft distance from the Sun is well-matched by an expected $1/r$ dependence, but with an added constant brightness of ~40 Rayleighs. This additional brightness is a possible signature of the hydrogen wall at the heliopause or of a more distant background. Ongoing observations are planned at a cadence of roughly twice per year.

**Plain Language Summary**

Long-term observations made with the Alice instrument on the New Horizons spacecraft confirm measurements made 30 thirty years earlier with the Voyager spacecraft. Both sets of data are best explained if the observed ultraviolet light is not only a result of the scattering of sunlight by hydrogen atoms within the solar system, but includes a substantial contribution from a distant source. This distant source could be the signature of a "wall" of hydrogen, formed near where the interstellar wind encounters the solar wind, or could be more distant. Similar future observations from New Horizons are planned about twice each year.

# 1 Introduction

The interplanetary medium (IPM) has been investigated by measuring sky Lyman-α (Lyα) emissions of atomic hydrogen (at a wavelength of $\lambda$=121.6 nm) for decades [e.g., *Bertaux & Blamont*, 1971; *Fahr*, 1974; *Adams & Frisch*, 1977; *Holzer*, 1977; *Thomas*, 1978; *Ajello et al.*, 1994; *Murthy et al.*, 1999; *Quémerais et al.*, 2009; *Izmodenov*, 2009; *Quémerais et al.*, 2010; *Izmodenov et al.*, 2013; *Bzowski et al.*, 2013; *Quémerais et al.*, 2013; *Katushkina et al.*, 2017]. These emissions arise when solar Lyα photons are scattered by neutral interstellar hydrogen atoms as they travel through the solar system [e.g., *Brandt & Chamberlain*, 1959; *Meier*, 1977; *Keller et al.*, 1981; *Braskén & Kyrölä*, 1998; *Quémerais & Bertaux*, 1993; *Izmodenov et al.*, 2013; *Bzowski et al.*, 2013; *Quémerais et al.*, 2013].

Near Earth, the brightness of IPM Lyα varies with the solar cycle and direction on the sky (both by about a factor of two), but is typically about 700 Rayleighs (i.e., about $7 \times 10^8$ photons cm$^{-2}$ s$^{-1}$ over the entire sky) [*Ajello et al.*, 1987]. For comparison, the direct Lyα flux from the Sun at Earth is ~4-6 × 10$^{11}$ photons cm$^{-2}$ s$^{-1}$. The direct solar Lyα flux is much more important for inner solar system photochemistry, but the IPM brightness falls off more slowly than the direct flux [*Ajello et al.*, 1987], and the two sources are of roughly equal strength at the orbit of Neptune [*Broadfoot et al.*, 1989]. At larger distances, e.g., in the Kuiper belt and beyond, the IPM Lyα source dominates.

The interstellar wind of H atoms comes from an ecliptic longitude and latitude $(\lambda, \beta)$ = (252.5˚, 8.9˚) [e.g., *Lallement et al.*, 2010], about 50˚ from the apex of the Sun's way, at a velocity of $v_H$ ~ 20 km/s [e.g., *Quémerais et al.*, 2009]. For comparison, *McComas et al.* [2015] recently determined an upstream direction of $(\lambda, \beta)$ = (255.7˚, 5.1˚) for the interstellar wind of He atoms.

Solar radiation pressure (from the Lyα line) on the H atoms approximately balances the force of gravity. Since both forces vary as $1/r^2$, where $r$ is the distance from the Sun, H atom trajectories are nearly undeflected (unlike the paths of interstellar He atoms, which are focused downstream of the Sun) [e.g., *Bzowski et al.*, 2013]. The interstellar H atoms are mainly lost due to charge exchange with much faster solar wind protons. This H lifetime against charge exchange is ~2-4 $[r \, (\mathrm{AU})]^2$ weeks (where $r$ is the distance from the Sun in astronomical units), compared to a lifetime against photoionization of ~14-20 $[r \, (\mathrm{AU})]^2$ weeks [cf., *Bzowski et al.*, 2013], so that near the Sun a density cavity forms (a typical interstellar H atom travels ~4 AU per year). This density cavity modifies the brightness distribution of the IPM Lyα emissions in the inner solar system; near the Earth, the IPM Lyα brightness is about twice as large looking in the upwind direction as in looking downwind. Viewed from the outer solar system, however, the region near the Sun is always much brighter than any other direction. Moving outward from the Sun, since the scattering rate of solar Lyα emission falls off roughly as $1/r^2$ (as long as the scatterers are of uniform density and are optically thin), then the IPM brightness looking in the direction away from the Sun should fall off roughly as $1/r$, since it is proportional to the integral of the scattering rate outward from $r$.

At a interplanetary atomic hydrogen number density of $n_H$ ~ 0.1 cm$^{-3}$ and temperature of $T_H$ ~ 15,000 K [e.g., *Izmodenov et al.*, 2013], the path length for unit optical depth is ~14 AU at line center, and at a Doppler width away from line center the path length for unit optical depth is ~38 AU. The solar Lyα line is very broad (with a full-width at half maximum, FWHM ~ 0.1 nm ≈ 250 km/s) compared to the width (16 km/s or 0.057 nm) and offset (from ~ -20 km/s to ~ +20 km/s) of the emissions scattered from the IPM hydrogen, so that the illumination can be taken to be approximately the same throughout the solar system, after accounting for extinction [e.g., *Wu & Judge*, 1979].

## 2 Observations

The New Horizons mission, which carries an ultraviolet spectrograph and which explored the Pluto system in July 2015 [*Stern et al.*, 2015] is currently en route to Kuiper Belt Object (KBO) 2014 MU69 (aka "Ultima Thule"), provides the first platform for observations of IPM Lyα emissions in the outer solar system since the Voyager spacecraft. During a few of its annual checkouts (ACOs) on the cruise to Pluto the Alice ultraviolet spectrograph [*Stern et al.*, 2008] was used to observe the IPM Lyα signal along a fixed great circle on the sky [*Gladstone, Stern, & Pryor*, 2013]. During ACOs 1, 2, and 4, great-circle swaths were observed at 90˚ from an ecliptic coordinates apex direction $(\lambda, \beta)$ = (51.3˚, 44.8˚), which avoided some bright UV stars and the Sun while passing within ~33˚ of the IPM downstream and upstream directions. During the Pluto encounter, the single great circle observation was upgraded to six evenly spaced great circles, allowing a sort of "all-sky" IPM Lyα map, with ecliptic coordinate apexes spaced every ~30˚ in ecliptic longitude $(\lambda_1, \beta_1)$ = (89.3˚, -4.9˚), $(\lambda_2, \beta_2)$ = (59.4˚, -10.1˚), $(\lambda_3, \beta_3)$ = (29.0˚, -12.4˚), $(\lambda_4, \beta_4)$ = (358.3˚, -11.4˚), $(\lambda_5, \beta_5)$ = (328.3˚, -7.4˚), and $(\lambda_6, \beta_6)$ = (298.7˚, -1.3˚). Figure 1 indicates where these different observations were performed, projected onto the plane of the ecliptic, as the New Horizons spacecraft follows its trajectory past the Pluto system toward 2014 MU69 and out into the Milky Way galaxy.

The Alice ultraviolet spectrograph on NH is comprised of a telescope section, a Rowland-circle-design spectrograph, a double-delay-line (DDL) microchannel plate (MCP) detector, and associated electronics and mechanisms [*Stern et al.*, 2008]. The entrance slit has two sections: a 2˚ × 2˚ "Box" and a 0.1˚ × 4˚ "Slot". The spectral bandpass is 52-187 nm, sampled by ~745 0.18-nm spectral pixels, and the filled-slot spectral resolution is ~0.9 nm. The Alice effective area has been determined through regular observations of UV-bright stars, whose absolute fluxes are fairly well known from IUE measurements. Consideration of where the filled-slit IPM Lyα emissions fall on the bare, KBr-coated, and CsI-coated sections of the Alice MCP leads to an overall sensitivity of the New Horizons Alice spectrograph to IPM Lyα of $S_{NH}$ ~ 5.5 counts/s/R [*Gladstone, Stern, & Pryor*, 2013; additional Alice calibration information is provided by *Greathouse et al.* 2010 and *Young et al.* 2018]. In comparison, for the Voyager 1 ultraviolet spectrometer the IPM Lyα sensitivity is $S_{V1}$ ~ 0.011 counts/s/R. The Alice spectrograph is off most of the time, and the total charge extraction in the region of the detector illuminated by Lyα is measured to be ~0.01 C/cm$^2$; this is large enough that gain sag has likely begun, but small enough that the sensitivity is unchanged.

The circumstances during the observations presented here are provided in Table 1. Most of the IPM Lyα observations made during the New Horizons mission have been different in some way. In ACO-1 a voltage ramp-up delay left Alice in histogram mode, and histogram exposures ($t_{EXP}$ = 30 s each) were gathered over only about one-half of the planned great circle. During ACO-2, pixel list data were obtained over two complete spacecraft spins of one hour each in duration. For ACO-4, histogram data ($t_{EXP}$ = 80 s each) were obtained over a single spacecraft spin. For the Pluto system flyby, 6 great circle count-rate-only observations were made during P-28 and P+1 days (relative to the closest approach of New Horizons to Pluto, at 11:48:30 UTC on 14 July 2015), at a scan rate of 1˚/s, with occasional aperture door closings to avoid observing several exceptionally bright stars (all other star crossings remain in the data, most of which are near the galactic plane, which is easily discernable in Fig. 2). For HEL-1 and HEL-2, the same 6 great circle count-rate-only scans were made, but at a slower scan rate of 0.1˚/s, and with near downstream and upstream histogram exposures of 300s and 3600s each to provide high-quality spectra. The directions for the near downstream and upstream histogram spectra were chosen to be close to both the interstellar wind direction and the galactic poles, while also being free of UV-bright stars. The observation strategy begun with HEL-1 and HEL-2 is planned to continue as long as possible.

Figure 2 shows the most recent great circle swaths, obtained during HEL-2 observation on 19 September 2017, overlain on a model map [*Pryor et al*., 2008; 2013] in ecliptic coordinates. The swath brightnesses shown in Figure 2 are based on count rates from housekeeping data, with a small correction for detector dead time, the subtraction of ~120 counts/s background (~100 counts/s due to particles from the radioisotope thermoelectric generator (RTG) plus ~20 counts/s due to fiducial "stim" counts), and conversion into Rayleighs assuming the filled-slit IPM sensitivity of $S_{NH}$ ~ 5.5 counts/s/R that was estimated above. The IPM Lyα brightness is largest in the direction closest to the Sun, and the ratio of the brightest/dimmest signal is ~4. The single scattering *Pryor et al*. [2008; 2013] IPM model generally underestimates the observed IPM Lyα brightness, and has been scaled up by a factor of 1.8× to agree more closely with the data. This scaling results in an overestimate in directions near the Sun but provides reasonable agreement in other directions. It is expected that multiple scattering effects become increasingly important as *r* increases, and this may partly account for the disagreement between model and data. Figure 3 presents the data mapped in Fig. 2 as they were observed on 19 September 2017, along with the expected brightness from the model and the ecliptic longitude and latitude pointing of the center of the Alice "Box".

Using the Alice histogram data from the three ACO and two HEL observations (i.e., spectral images), a search for IPM emissions other than Lyα was made. Figure 4 shows spectra integrated over the swaths from ACO-1, ACO-2, and ACO-4, with bright star contamination removed, plus the 3600s near-upstream spectra from the HEL-1 and HEL-2 observations. Most of the spectral structure is a result of the broad wings of the Lyα line and detector edge effects (at short and long wavelength limits). A feature at 58.4 nm due to interplanetary helium scattering the corresponding solar line is not clearly seen in Fig. 4, as it is near the short wavelength end of the Alice bandpass. This non-detection is likely a combination of: 1) detector edge effects; 2) the RTG background (~100 counts/s over the entire detector, amounting to ~2 count/s over the 58.4 nm line; and 3) the relatively low sensitivity at this wavelength ($S_{584} \sim 0.03$ counts/s/R).

However, the Lyβ line at 102.6 nm is clearly seen in all five spectra, and likely results from resonant scattering of the solar Lyβ line by interplanetary H atoms ($S_{Ly\beta} \sim 0.10$ counts/s/R). This feature is quite faint, at ~0.5-1 Rayleigh, as seen in Table 1 (the IPM brightnesses and Lyα / Lyβ brightness ratios in Table 1 are derived from the Figure 4 spectra). The IPM Lyα / Lyβ brightness ratios from the ACO observations agree fairly well with the Voyager result (from $r$<15 AU observations) of 700±200 [*Murthy et al*., 1999] and with a measurement of 634 by the extreme ultraviolet spectrometer on Galileo [*Hord et al*., 1991]. However, the ratio decreases further from the Sun, probably as multiple scattering of Lyα becomes more important. As seen in Table 1, the IPM Lyα / Lyβ ratio is larger than the solar Lyα / Lyβ ratio, probably because of the lack of substantial multiple scattering of Lyβ in the IPM, due to its substantial likelihood of a branching to the Hα transition whenever it is scattered by an H atom.

As mentioned above, while direct solar Lyα falls off as $1/r^2$, IPM-scattered solar Lyα (viewed away from the Sun) is expected to fall off as $1/r$, so that it becomes more important in the outer solar system (e.g., for photochemistry in the upper atmospheres of Neptune, Triton, and Pluto, especially on the night or winter hemispheres of these bodies; for radiolysis of the surfaces of KBOs, etc.). In the early solar nebula this kind of photochemistry by IPM Lyα was probably significant throughout the solar system [*Gladstone*, 1993; *Throop*, 2011]. The initial Alice ACO data suggested that the falloff was even more gradual with distance from the Sun - closer to ~$1/r^{1/2}$ [*Gladstone, Stern, & Pryor*, 2013]. Figure 5 compares how the ACO and more recent Alice data compare with the *Hall* [1992] Voyager data (scaled downward by 2.4× as recommended by *Quémerais et al.* [2013] – note that this has been challenged by *Ben-Jaffel & Holberg* [2016]). The agreement between the two data sets, separated in time by ~30 years, is very good. Furthermore, as originally suggested by *Hall* [1992] and more recently explored by *Katushkina et al.* [2017], it appears that a more plausible explanation for the gradual falloff in IPM Lyα brightness viewed in the upwind direction is that the expected $1/r$ falloff is accompanied by a constant background emission, from beyond the heliopause. While a possible source for this added background is scattered solar Lyα from a "hydrogen wall" just outside the heliopause [*Baranov*, 1990; *Baranov & Malama*, 1993], the analysis of Voyager 1 UVS data at 90-130 AU by *Katushkina et al.* [2017] strongly suggests that the source is more distant and is not consistent with reflected solar Lyα emissions. Future observations by New Horizons may enable more complete understanding of this background emission.

## 3 Conclusions

The series of IPM Lyα observations being made by the New Horizons Alice ultraviolet spectrograph represent the first such observations from outside of Saturn's orbit since Voyager. The key result so far is that, using the scaling of the Voyager UVS data suggested by *Quémerais et al.* [2013], there is very good agreement with the results of *Hall* [1992] regarding the unexpectedly slow decline in the IPM Lyα brightness seen in the upstream direction. This result strongly suggests the presence of a ~40 R background of Lyα from beyond the heliopause, similar to the 25 R background recently found by *Katushkina et al.* [2017] in their analysis of 90-130 AU Voyager 1 UVS data.

With a much higher sensitivity than Voyager UVS, the Alice spectrograph continues to investigate the distribution and brightness of interplanetary Lyα emissions in the outer solar system, which should lead to a much improved understanding of the structure of the heliopause. Estimates of future RTG performance suggest that Alice could be operable for ~15-20 or so more years (out to ~90 AU or farther), although other factors are also important (e.g., how much fuel remains after the 2014 MU69 flyby). Further IPM Lyα observations are planned, on a more or less twice-per-year basis, for as long as possible.

### Acknowledgments, Samples, and Data

We thank the NASA's New Horizons project for their excellent support. We thank Eric Quémerais for valuable discussions and the reviewers for helpful comments. New Horizons is funded by NASA, whose financial support we gratefully acknowledge. The Alice data used in this work is available (Cruise and Pluto flyby data) or will be available (post-flyby heliospheric data) from the Small Bodies Node of NASA's Planetary Data System (e.g., https://pds-smallbodies.astro.umd.edu) under the New Horizons section of the Missions subdirectory.

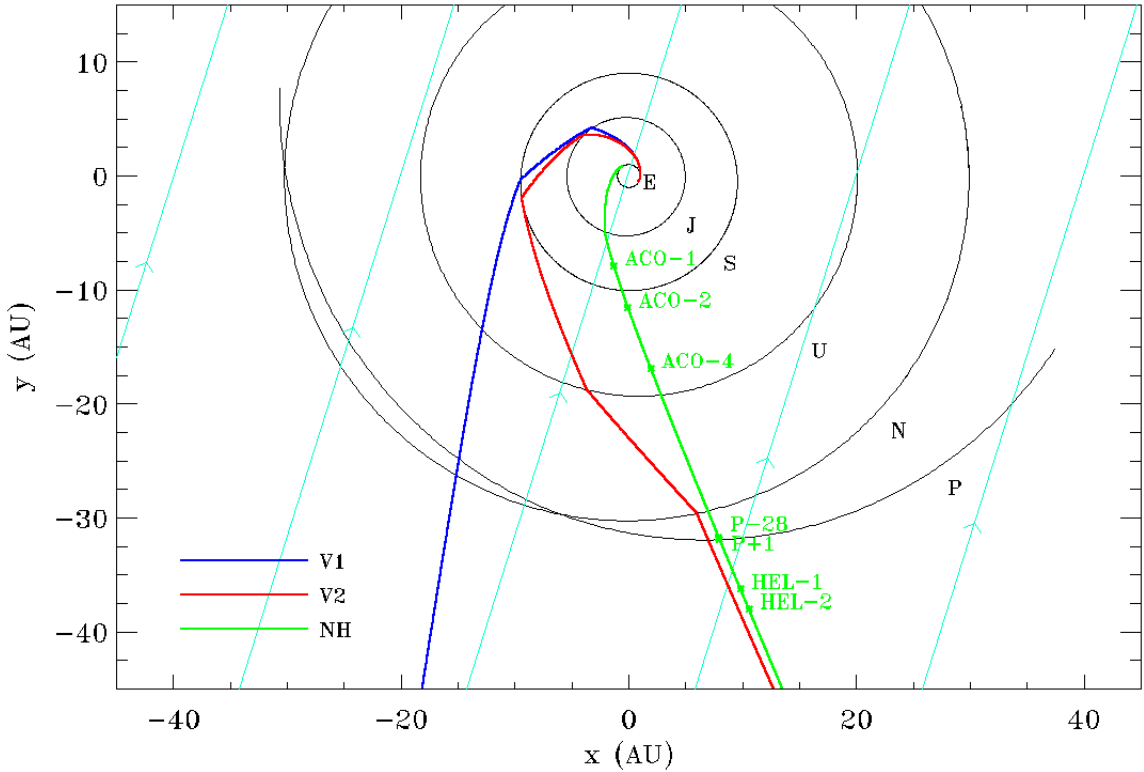

**Figure 1.** The trajectory of New Horizons (green) is shown projected onto the plane of the ecliptic, with the trajectories of Voyager 1 (blue) and Voyager 2 (red) for comparison. The projected interstellar hydrogen wind direction is indicated (light blue). New Horizons travels approximately radially upwind at a rate of ~3 AU/year, and is within 2° of the ecliptic. The New Horizons Alice ultraviolet spectrograph has been used to occasionally observe interplanetary medium (IPM) Lyα along great circles on the sky at certain points during the mission: 1) during some cruise-phase annual checkouts (ACO-1, ACO-2, and ACO-4); 2) the Pluto system flyby (P-28, P+1); and 3) post-flyby heliophysics observations (HEL-1, HEL-2). After the upcoming flyby of Kuiper Belt Object (KBO) 2014 MU69 (aka "Ultima Thule") on 1 January 2019, these IPM observations are planned to continue at approximately 6-month intervals for as long as possible.

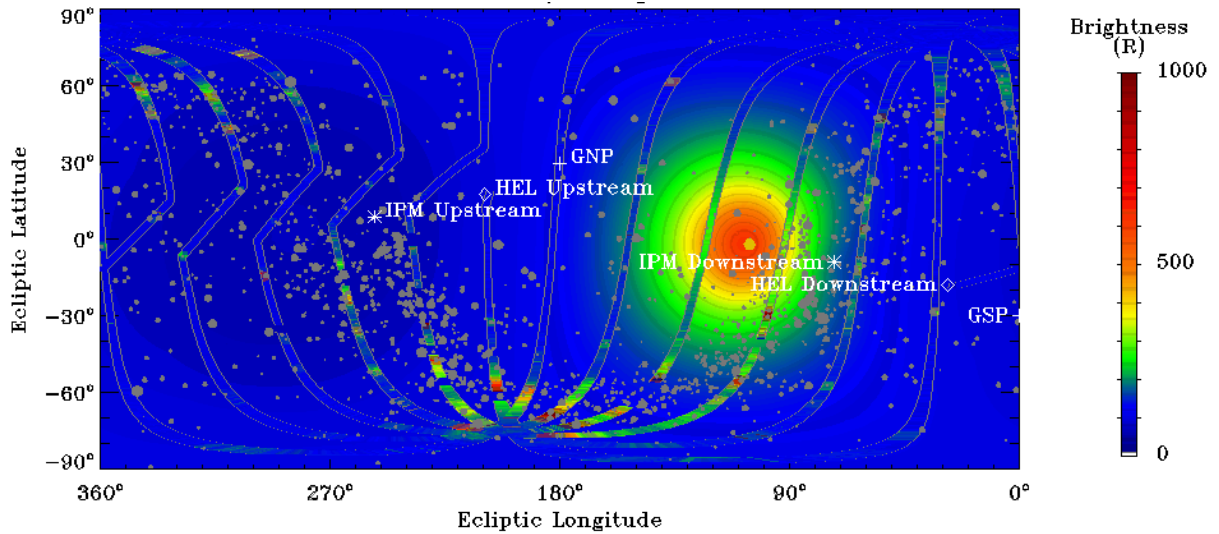

**Figure 2.** Interplanetary medium (IPM) Lyα HEL-2 measurements by the Alice spectrograph overlain as six great-circle swaths (outlined in gray) on a model IPM brightness map (in ecliptic coordinates) appropriate for the HEL-2 location of the New Horizons spacecraft (at $r_{NH} = 39.5$ AU on 19 September 2017), with a color bar linear in brightness units of Rayleighs. Bright UV stars (gray circles, sized according to brightness), mostly in the galactic plane, account for most of the small-scale bright features. Data were also collected during slews between the great circles (e.g., seen in the -10˚ to 30˚ latitude range on the left side of the figure). The model has been scaled upward by a factor of 1.8 to better agree with the data. The Sun (yellow circle) and the downstream and upstream directions for interstellar H atoms [*Lallement et al.*, 2010] are indicated, along with the galactic poles and the directions of the 60-minute spectra taken during the HEL series of observations made during the extended mission of New Horizons.

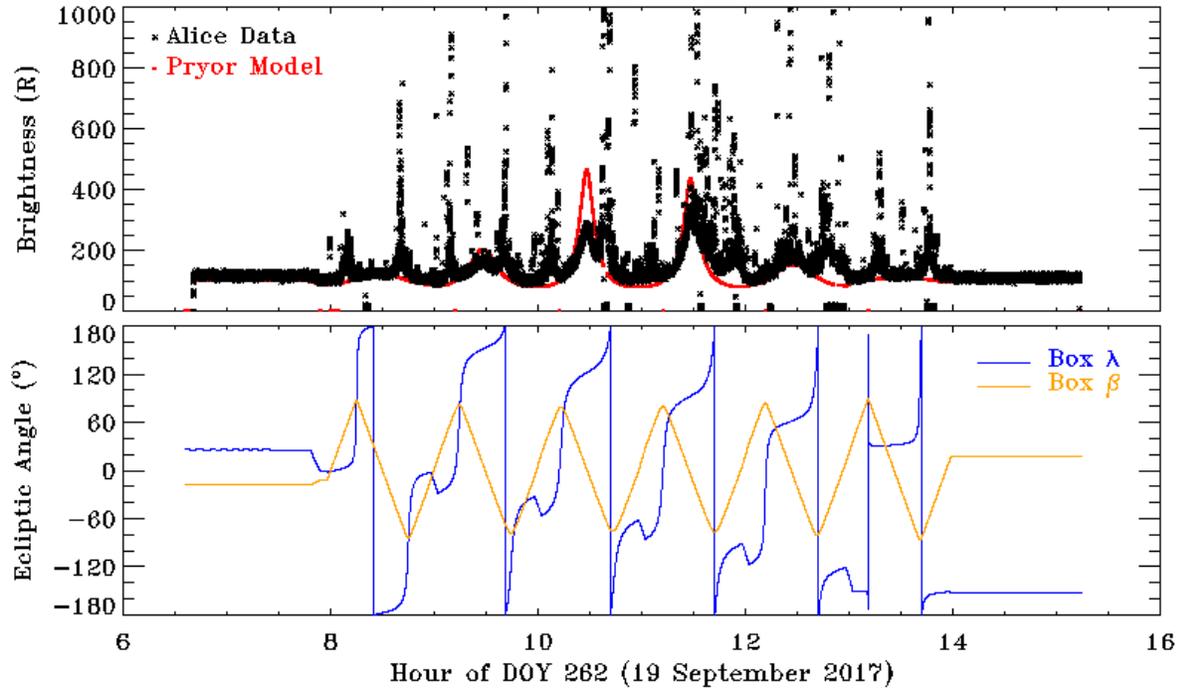

**Figure 3.** Interplanetary medium (IPM) Lyα HEL-2 brightnesses observed by the Alice spectrograph (top panel, black asterisks; as shown in Figure 2) during the 6-great-circle scan are plotted as a function of time, along with model IPM brightnesses (scaled by 1.8×; top panel, red asterisks) and ecliptic longitudes and latitudes for the pointing direction of the center of the Alice "Box" (bottom panel, blue and orange lines, respectively). Note that most of the "spikes" in brightness are due to UV-bright stars passing through the Alice slit.

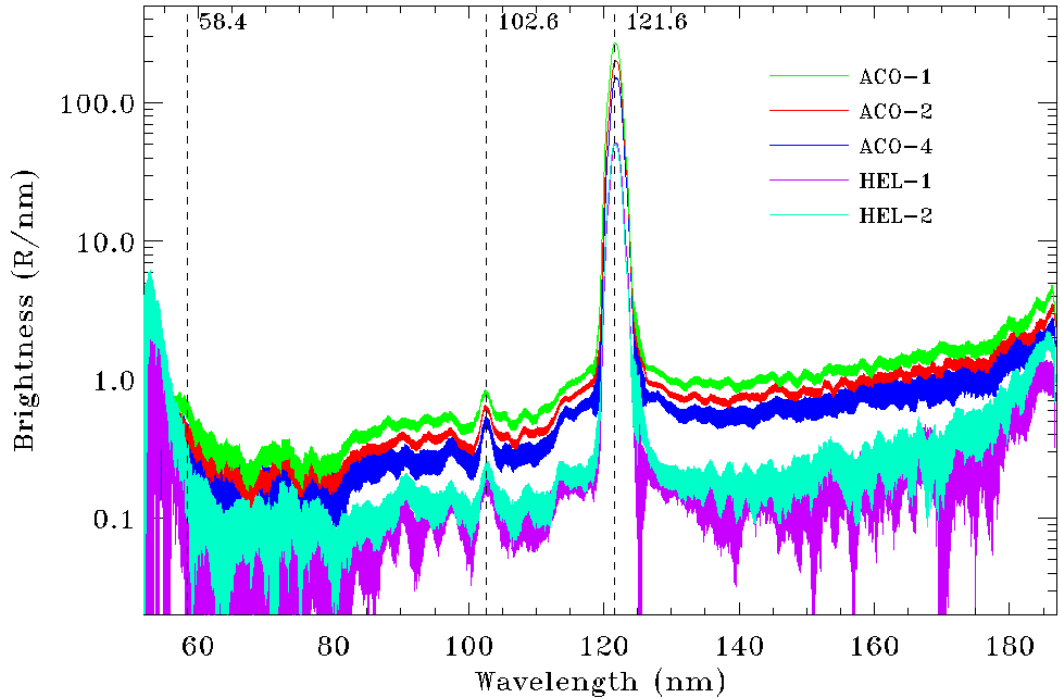

**Figure 4.** Interplanetary medium spectra (±1σ) observed by the New Horizons Alice ultraviolet spectrograph during ACO-1, ACO-2, ACO-4 (averaged over their great-circle swaths, with star contributions removed), and during the HEL-1 and HEL-2 upstream-direction 60-minute histograms, using only detector rows corresponding to the 0.1˚-wide "slot" part of the slit (for improved spectral resolution). Besides Lyα, the similarly-excited Lyβ line at 102.5 nm is detected, although another expected resonance line of He atoms at 58.4 nm is not clearly seen. For comparison, the galactic background at FUV wavelengths is estimated to be <13 mR/nm [*Henry et al.*, 2015].

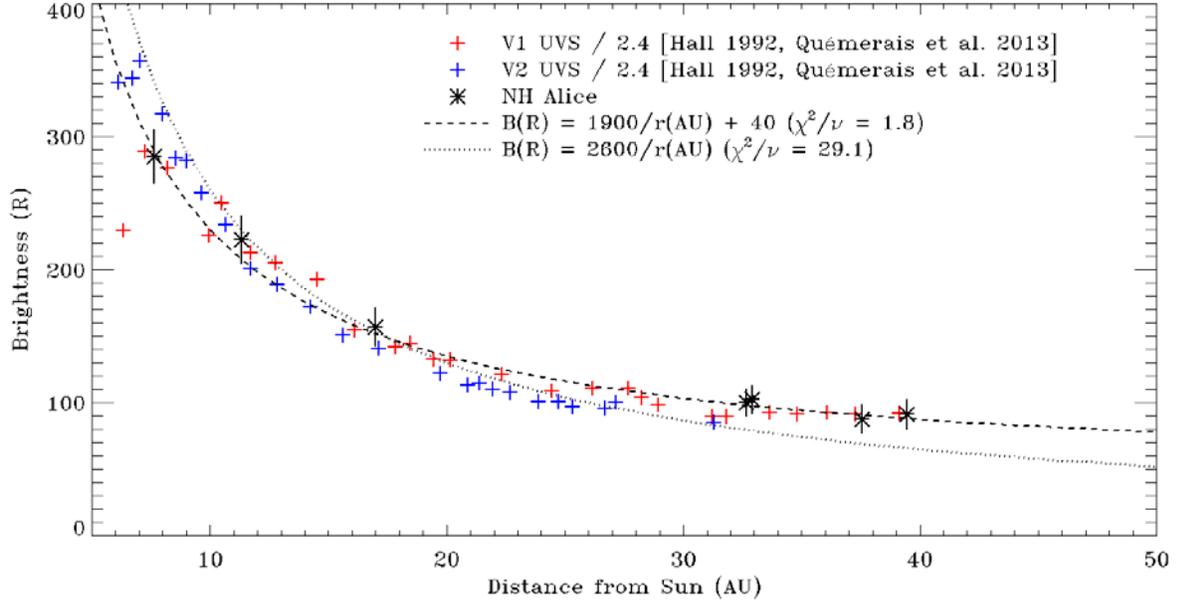

**Figure 5.** Observed falloff in brightness of IPM Lyα emission viewed in the upwind direction measured by the UVS on Voyager 1 (red crosses) and Voyager 2 (blue crosses) [*Hall* 1992], scaled downward by 2.4× as recommended by *Quémerais* [2013], and by New Horizons Alice (black asterisks, with 3-σ error bars). All data have also been scaled by a factor of $(3 \times 10^{11}$ photons/cm$^2$/s) / $\pi F_{Sun}$, where $\pi F_{Sun}$ is the estimated sub-spacecraft 6-day average solar Lyα flux at 1 AU (e.g., as provided in Table 1 for New Horizons data). Two simple empirical models for the Alice data are shown for comparison (with reduced $\chi^2$ indicated), both with the expected $1/r$ brightness dependence, but one with an additional distant upstream brightness of 40 R.

| Table 1. Circumstances of New Horizons Alice Interplanetary Medium Lyα Observations | | | | | | | |
|---|---|---|---|---|---|---|---|
| **Observation Name** | **ACO-1** | **ACO-2** | **ACO-4** | **P-28** | **P+1** | **HEL-1** | **HEL-2** |
| **Date** | 10/07/07 | 10/18/08 | 06/19/10 | 06/16/15 | 07/15/15 | 01/28/17 | 09/19/17 |
| **Start UTC** | 02:13:45 | 11:47:51 | 09:07:42 | 19:34:05 | 19:01:04 | 11:55:30 | 06:35:41 |
| **End UTC** | 03:10:44 | 13:51:02 | 11:21:46 | 20:24:31 | 19:51:31 | 20:34:09 | 15:14:20 |
| **Scan Rate (°/s)** | 0.1 | 0.1 | 0.1 | 1.0 | 1.0 | 0.1 | 0.1 |
| $\mathbf{r_{NH}}$ **(AU)** | 7.624 | 11.337 | 16.991 | 32.679 | 32.919 | 37.56 | 39.47 |
| $\mathbf{\lambda_{NH}}$ **(°)** | 259.32 | 269.33 | 276.56 | 284.02 | 284.09 | 285.13 | 285.49 |
| $\mathbf{\beta_{NH}}$ **(°)** | 1.17 | 1.50 | 1.72 | 1.91 | 1.91 | 1.94 | 1.95 |
| **NH-Sun-Earth (°)**[†] | +113.96 | +115.96 | -8.93 | -18.99 | +8.88 | -156.56 | +70.77 |
| **Lyα $\pi F_{Sun}$ @1AU** **($10^{11}$ photons/cm²/s)**[‡] | 3.37 | 3.24 | 3.44 | 4.20 | 4.11 | 3.42 | 3.27 |
| **Lyβ $\pi F_{Sun}$ @1AU** **($10^{9}$ photons/cm²/s)**[‡] | 3.47 | 3.44 | 3.71 | 5.13 | 4.72 | 4.27 | 4.38 |
| **Solar Lyα/ Lyβ** | 97.2 | 94.0 | 92.7 | 81.8 | 87.0 | 80.3 | 74.7 |
| **IPM Lyα (R)**[*] | 547.4±1.9 | 404.3±1.2 | 300.6±1.3 | - | - | 100.0±0.3 | 99.3±0.8 |
| **IPM Lyβ (R)**[*] | 0.73±0.09 | 0.67±0.06 | 0.37±0.11 | - | - | 0.25±0.06 | 0.39±0.07 |
| **IPM Lyα/ Lyβ**[*] | 754±90 | 602±52 | 811±233 | - | - | 399±102 | 254±43 |

[†]Angles are negative from solar conjunction until opposition and positive from opposition until solar conjunction
[‡]Solar fluxes from TIMED/SEE L3A line irradiances, corrected for relative ecliptic longitudes of Pluto and Earth as seen from the Sun, taken from http://lasp.colorado.edu/lisird/data/timed_see_lines_l3a/
[*]Upstream spectra were not recorded during the P-28 and P+1 observations